\documentstyle[emulateapj]{article}

\input psfig

\def\beq{
\begin{equation}
}
\def\eeq{
\end{equation}
}
\def\simge{\mathrel{%
   \rlap{\raise 0.511ex \hbox{$>$}}{\lower 0.511ex
\hbox{$\sim$}}}}
\def\simle{\mathrel{
   \rlap{\raise 0.511ex \hbox{$<$}}{\lower 0.511ex
\hbox{$\sim$}}}}

\def\Real{{\rm I\mathchoice{\kern-0.70mm}{\kern-0.70mm}{\kern-0.65mm}%
  {\kern-0.50mm}R}}

\font \bolditalics = cmmib10
\def\bx#1{\leavevmode\thinspace\hbox{\vrule\vtop{\vbox{\hrule\kern1pt
        \hbox{\vphantom{\tt/}\thinspace{\bf#1}\thinspace}}
      \kern1pt\hrule}\vrule}\thinspace}

\def \vc #1{{\textfont1=\bolditalics \hbox{$\bf#1$}}}

\def\eg{{\bf e}}

\def\ggr{{\bf g}}

\def\nablag{{\vc \nabla}}
\def\xig{{\vc \xi}}

\def\thetag{{\vc \theta}}

\def\gammag{{\vc \gamma}}

\def\epsilong{{\vc \epsilon}}

\def\Vg{{\bf V}}

\def\Mg{{\bf M}}
\def\Pc{{\cal P}}

\def\Ac{{\cal A}}
\def\Cc{{\cal C}}

\def\Sc{{\cal S}}

\def\be{\begin{equation}}
\def\ee{\end{equation}}
\def\ba{\begin{eqnarray}}
\def\ea{\end{eqnarray}}

\def\d{{\rm d}}

\begin{document}

\title{Lensing effect on the relative orientation between
the Cosmic Microwave Background ellipticities and the distant galaxies}

\author{L. Van Waerbeke\altaffilmark{1}, F. Bernardeau\altaffilmark{2},
K. Benabed\altaffilmark{2}}

\altaffiltext{1}{CITA, 60 St George Str., Toronto, M5S 3H8, Canada}
\altaffiltext{2}{SPhT, C.E. de Saclay, F-91191 Gif-sur-Yvette Cedex, France}

\begin{abstract}
The low redshift structures of the Universe act as lenses
in a similar way on the Cosmic Microwave Background light and on the distant
galaxies (say at redshift about unity). As a consequence, the CMB
temperature distortions are expected to be statistically correlated  
with the galaxy shear, exhibiting a non-uniform 
distribution of the relative angle between the CMB and the
galactic ellipticities.  
Investigating this effect we find that its amplitude
is as high as a $10\%$ excess of alignement between CMB and the
galactic ellipticities relative to the uniform distribution.
The relatively high signal-to-noise ratio we found should makes
possible a detection with the planned CMB data sets,
provided that a galaxy survey follow up can be done on a sufficiently
large area. It would provide a complementary bias-independent
constraint on the cosmological parameters.

\end{abstract}


\section{Introduction}

The analysis of gravitational lensing of the primordial fluctuations
in the Cosmic Microwave Background (CMB) is of growing interest since
it gives a direct probe of the mass distribution up to very high
redshift. The detection of the lens effects on the CMB would be very
precious for constraining the cosmological parameters, with no
ambiguities about the distance of the source plane. Although the
effect on the CMB power spectrum is rather weak and affects only small
scales ($l>1000$) it is now recognized that gravitational lensing can
produce some specific features worth for investigation. In this paper
we focus our investigations on the temperature maps only although it
has been noticed recently that lenses can also  affect significantly
the polarization properties of the CMB maps (\cite{zs98a,bb99}).

So far there are two distinct effects that kept the attention:
the small (sub-arcmin) scale
strong deformation of the temperature fluctuations
which can be used to probe galaxy cluster
gravitational potential (\cite{zs98b}, \cite{ms})  and an intermediate
and large scale
statistical effect (Bernardeau 1997, 1998, \cite{sz99}) which is a way
to determine the cosmological parameters and provides a consistency
test against other CMB analysis. The detection of the small scale
lens effects, remains to be proved
feasible. In particular the accumulation of secondary anisotropies
(such as the Sunyaev-Zeldovich and Ostriker-Vishniac effects, and the
non-linear Sachs-Wolfe effect) will mask the primordial lens effects
and make the interpretation of lensed secondary anisotropies
dependent on the {\it unknown} redshift where the
secondary anisotropies where generated.  At scales above 1 or 2 arcmins,
the interpretation
of the lens effects should be more straightforward but the number density of
fluctuations is not large enough to allow an accurate mass reconstruction of
the lens (in particular in the case of clusters of galaxies). Therefore at
scales larger than a few arcmin only a statistical detection of the lens effect
seems possible. More specifically Bernardeau (1998) investigated the
effect of lensing  by  the large-scale structures on
the distribution of the CMB ellipticities.  The
consequences are the same as for the lensed
distant galaxies: the gravitational distortion induces an
excess of elongated structures of CMB ellipticities.
The intrinsic CMB ellipticity distribution being known for a
Gaussian field (e.g. Bond \& Efstathiou 1987) it is then
possible to compute the lensed distribution. The lensed
distribution is unfortunately rather close to the unlensed one,
in particular because the smoothing caused by the CMB beam
tends to circularizes the local structures.  The orientation of the
local ellipticity is expected to be much more robust against the
smoothing effects and therefore more efficient in tracing the lens
effects.  Due to the low number density of structures on CMB maps such
effects can be hardly detected in CMB maps alone, so in this paper we
rather investigate the possible cross-correlation of CMB ellipticities with
the distant galaxy ellipticities.

It has been recognized
before that a fair fraction of the lenses that act on CMB temperature
maps are at low redshift (\cite{SSS}). There are different consequences.
Temperature maps and galactic density survey should exhibit some
correlations (\cite{SSS}), and it induces a non-zero three-point
function (or equivalently a bispectrum) in the CMB maps (\cite{gs},
\cite{sg}). This is due to the coupling between lens effects and low
redshift primary or secondary anisotropies such as the integrated
Sachs-Wolfe effect or the Sunyaev-Zeldovich effect.

In this paper we focus our analysis in the expected 
correlation between the CMB ellipticities and those of
distant galaxies. We indeed expect their relative angle
to be not-uniformly distributed unlike what would
happend if there were no lensing effects. We examine here
the amplitude and the observability of this effect. 
It is worth stressing that, unlike in the analysis performed by 
\cite{SSS},
the correlation signal we are aiming at is independent on any
possible galaxy bias. It would then provide a complementary
constraint on the fundamental cosmological parameters a priori
independent on those provided by the observations of the primary anisotropies.

%As a consequence we expect that
%the relative orientations of the CMB ellipticities and those of
%distant galaxies to be not-uniformly distributed unlike what would
%happend if there were no lensing effects. This paper is an analysis of
%the amplitude and the observability of this effect.
%One aspect of the cross-correlation of the shear of the distant galaxies with
%the CMB anisotropies has been investigated by \cite{gs} and \cite{sg}.
%The authors searched for a non-Gaussian signature by calculating
%the amplitude of the cross-bispectrum between the shear and
%the most dominant secondary anisotropies. They found a signal-to-noise rather modest
%due to the fact that the measurement of bispectra is very noisy. Here we focuss
%on the CMB-galaxy ellipticity cross-correlation which has a non-vanishing contribution
%to the second order of the perturbation theory, which is in essence much more
%easy to detect than a third order contribution.

We discuss the physical mechanism, and introduce the relevant
quantities in Section 2.  In Section 3 we investigate the amplitude of
the effect for different cosmologies and we calculate the
signal-to-noise ratio in Section 4. Section 5 is a discussion on the
contribution of the different possible foreground contaminations before
we conclude in Section 6.

\section{Lensing of CMB ellipticities}

A bundle of light rays coming from a high redshift source from the
direction $\thetag=  (\theta_x,\theta_y)$ is deflected by a quantity
$\xig(\thetag)$, whose derivatives $\kappa=\nablag\cdot\xig$ and
$\gammag=(\partial_x \xi_x-\partial_y\xi_y;2\partial_x\xi_y)$ describe
the isotropic and anisotropic deformation of the light bundle. The
convergence $\kappa$ and the shear $\gammag$ depend on the integrated
gravitational potential along the line-of-sight $\thetag$ and distort
the image of distant galaxies as well as the CMB fluctuations.

At any position on the CMB temperature map, we can define an ellipticity
$\eg$ from the curvature of the temperature field $\delta_T$:
\begin{equation}
\eg=\left({\partial^2_x\delta_T-\partial^2_y\delta_T\over \partial^2_x\delta_T
+\partial^2_y\delta_T};{2\partial_{xy}\delta_T\over \partial^2_x\delta_T
+\partial^2_y\delta_T}\right).
\label{edef1}
\end{equation}
This relation is similar to the ellipticity of a galaxy defined from
its second order moments.  A peak of temperature with the same
curvature on both axis has a zero ellipticity, but in opposition with
the galaxies, the CMB ellipticity can take any value between zero
(circular peaks) and infinity (symmetric saddle points). However, it
is always meaningful to define the orientation of the CMB ellipticity
$\theta_{\eg}=\arctan(e_2/e_1)$, which runs from 0 to $2\pi$. The
gravitational lensing effect tends to stretch the structures and
therefore to produce an excess of elongated structures relative to the
number of rather round objects (Bernardeau (1998)). The lenses tend
also  to align the CMB ellipticity with the shear $\gammag(\thetag)$
acting on the CMB at the angular position $\thetag$. This is similar
to the effect which occurs on the ellipticity of distant galaxies,
althought the corresponding shear
$\gammag_g(\thetag)$ cannot be identified with $\gammag(\thetag)$
since the galaxies are at much lower redshift. The variables
$\gammag(\thetag)$ and $\gammag_g(\thetag)$ are however correlated
because the light coming from either the CMB or the distant galaxies
are passing through the same portion of low-redshift Universe, and
consequently, for a given line-of-sight $\thetag$, the CMB
ellipticities are preferentially aligned with the distant galaxies.

In the paradigm of inflation, the CMB fluctuations are Gaussian.
Therefore the un-lensed CMB ellipticity distribution is very specific
and furthermore independent on the shape of the temperature power
spectrum (\cite{be87}). The effect of gravitational lensing on the
statistic of a Gaussian random field can be calculated analytically, at
least using perturbative methods. According to (\ref{edef1}), the fields
of interest are the second derivatives of the temperature field, which
defines the CMB ellipticity. It is usefull to introduce the  matrix of
the second order derivatives $\Cc$
\begin{equation}
\Cc=\pmatrix{\partial_x^2 \delta_T & \partial_{xy} \delta_T \cr
\partial_{xy} \delta_T & \partial_y^2 \delta_T}=\pmatrix{\tau+g_1 &
g_2 \cr g_2 & \tau-g_1},
\label{curvmat}
\end{equation}
where the CMB ellipticity (\ref{edef1}) is defined as in Bond \& Efstathiou (1987)
as,
\begin{equation}
\eg={\ggr\over 2 \tau}.
\label{e_def}
\end{equation}
We want to calculate the effect of lensing on $\eg$ and write the
lensed ellipticity $\hat\eg$ as a function of $\eg$, the shear
$\gammag$ and the convergence $\kappa$. For this we need to calculate the
lensed quantities $\hat\ggr$ and $\hat\tau$ and expand the expressions
using the weak lensing approximation $(\gammag,\kappa)\ll1$.

The effect of weak lensing is only a re-mapping of the temperature
fluctuations through the displacement field $\xig(\thetag)$, with no
modification of the temperature amplitude,
\begin{equation}
\delta_T^{\rm obs}(\thetag)=\delta_T^{\rm prim}(\thetag+\xig(\thetag)).
\label{mapping}
\end{equation}
The magnification matrix is the Jacobian of the transformation between
the source and the image plane. It is given by the first derivatives
of the displacement field:

\begin{equation}
\Ac_{ij}=\delta^K_{ij}+\xi_{i,j}=\pmatrix{1-\kappa-\gamma_1 &
-\gamma_2 \cr -\gamma_2 & 1-\kappa+\gamma_1},
\label{amplifmat}
\end{equation}
where $\kappa$ is the convergence and $\gammag$ the shear. The lensed
curvature matrix calculated from Eq.(\ref{curvmat}) and
(\ref{mapping}) is (\cite{b98}):

\begin{equation}
\Cc^{\rm obs}_{ij}=\Ac_{ik}\Cc^{\rm
prim}_{kl}\Ac_{lj}+(\partial_k\delta_T)(\partial_j \xi_{k,i})
\label{LensShapeEq}
\end{equation}
The first term is identical to the lensing term for galaxies, and the
second term makes intervene the spatial derivatives of $\kappa$ and
$\gamma$. This second  term accounts for the variation of the
amplification matrix accross the CMB patterns. It cannot be a priori 
neglected because the scale at which the shear is estimated
corresponds roughly to the filtering scale, a scale  over which the shear might
significantly change. Note that for the
galaxy field, the shear is measured at a much smaller scale (the
galaxy size), and the shear gradient term is always neglected.
However we will assume later that the shear field is Gaussian (which
is a reasonable approximation for scales larger than a few arcmin), for
which the shear gradient is uncorrelated to the shear
itself. Therefore we expect the shear
gradients to play no role in the cross-correlation pattern. As a
result, in the following, the second term of Eq. (\ref{LensShapeEq})
will be ignored for the lensed CMB ellipticities as well.

It is
then easy to obtain the transformation equations for $\tau$ and $\ggr$:
\begin{eqnarray}
\hat\tau&\simeq&\tau(1-2\kappa)+2\gammag\cdot\ggr\nonumber\\
\hat\ggr&\simeq&\ggr(1-2\kappa)+2\tau \gammag.
\label{gtau_lensed}
\end{eqnarray}
The lensed ellipticity $\hat\eg$ is obtained from Eq.(\ref{e_def}) and
(\ref{gtau_lensed}):
\begin{equation}
\hat\eg\simeq\eg+4(\gammag\cdot\eg)\eg-\gammag
\label{e_lensed}
\end{equation}
Note that this is very similar to the galaxy ellipticity
transformation relations (\cite{ss95}) for which we have
$\hat\epsilong\simeq\epsilong-(\gammag\cdot\epsilong)\epsilong
+\gammag$ if $\epsilong$ is the intrinsic galaxy ellipticity. It is
remarkable that the lensed ellipticity does not depend on the
convergence $\kappa$.

\section{Galaxy-CMB ellipticities orientation}

Let $\theta$ be the angle between $\hat\eg$ and the shear
$\gammag(\thetag)$ at the redshift of the CMB, and $\theta_g$ the
angle between $\hat\eg$ and the shear of the distant galaxies
$\gammag_g(\thetag)$:
\begin{equation}
\cos(\theta)={\hat\eg \cdot \gammag\over \hat e \gamma} \ ; \\
\cos(\theta_g)={\hat\eg \cdot \gammag_g\over \hat e \gamma_g}
\end{equation}
We should emphasize that the angles $\theta,\theta_g$ are twice the
physical angles measured from the temperature map and the distant
galaxies (e.g.  $\theta_g=\pi$ means that a lensed temperature fluctuation
is perpendicular to a sheared distant galaxy, if they were both
intrinsically circular).

We want to calculate the probability distribution function
$\Pc(\theta_g)$ of the relative orientation between the lensed CMB
ellipticity and the sheared distant galaxies.  This can be done by
marginalizing the joint probability $\Pc(\hat\eg,\gamma,\gamma_g,
\theta,\theta_g)$ over $(\hat\eg,\gamma,\gamma_g,\theta)$.
We adopt the point of
view of conditional probability and write $\Pc(\hat
\eg,\gamma,\gamma_g,\theta,\theta_g)$ as,
\begin{equation}
\Pc(\hat \eg,\gamma,\gamma_g,\theta,\theta_g)= \Pc(\hat
\eg|\gammag)\Pc_{\rm lens}(\gamma,\gamma_g,\theta,\theta_g),
\label{desired_proba}
\end{equation}
where $\Pc_{\rm lens}(\gamma,\gamma_g,\theta,\theta_g)$ (which is a
pure lensing contribution), is the joint probability that the shear 
is $\gammag$ at the
last-scattering surface and $\gammag_g$ for the distant
galaxies. $\Pc(\hat \eg|\gammag)$ is the probability to observe a lensed
CMB ellipticity $\hat \eg$ when the shear is $\gammag$ at the
last-scattering surface. It is calculated from the ellipticity
distribution function $p(e)$ of the unlensed CMB (\cite{be87}):

\begin{equation}
p(e)\d e=p(e_1,e_2)e\d e={8e\over (1+8e^2)^{3/2}}\d e.
\end{equation}
{}From the Eq.(\ref{e_lensed}) and the weak lensing approximation
($\gammag \ll1$) we get,
\begin{eqnarray}
P(\hat \eg|\gammag)&=& P(\hat e_1,\hat e_2|\gammag)\hat e=
p(e_1,e_2)\left|{\partial e_i\over \partial \hat e_j}\right| \hat e
\nonumber\\ &\simeq& {8\hat e\over (1+8\hat e^2)^{3/2}}
\left(1-36{\hat\eg\cdot\gammag\over(1+8\hat e^2)}\right)
\end{eqnarray}
The integration over $\hat e$ gives the desired probability
(\ref{desired_proba}) as a function of $\Pc_{\rm lens}$ only, 
\begin{equation}
\Pc(\gamma,\gamma_g,\theta,\theta_g)=(1-3\sqrt{2}
\gamma\cos(\theta))\Pc_{\rm lens}(\gamma,\gamma_g,\theta,\theta_g). 
\label{proba_lens}
\end{equation}

\begin{figure}[t!]
\centering \psfig{figure=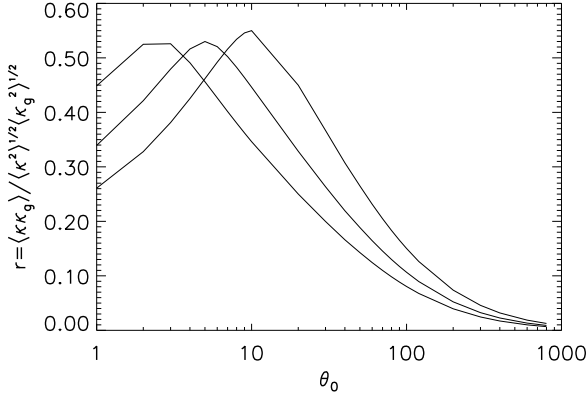,width=0.45\textwidth}
\caption{Correlation coefficient $r=\langle
\kappa\kappa_g\rangle/(\sigma\sigma_g)$ as defined in (\ref{coreq})
for LCDM model for a galaxy source plane at $z=1$. $r$ is given
as a function of the smoothing angle $\theta_0$ for $\kappa_g$.
The three curves from left to right correspond to a fixed beam
size $\theta_0^{CMB}$ of $2.5'$, $5'$ and $10'$. 
The stronger correlation is obtained when $\theta_0$ is comparable to the CMB
beam size.
\label{corcoef}}
\end{figure}

At this stage we need a model for
$\Pc_{\rm lens}(\gamma,\gamma_g,\theta,\theta_g)$. Both the observables
$\hat\eg$ and $\gammag_g$ will respectively be measured on the
temperature fluctuation map and a galaxy survey. Due to the limited beam size
of the bolometers, we cannot hope to measure $\hat\eg$ with a
resolution better than a few arcmin. Let $\theta_0^{\rm CMB}$ be this
beam size, and $\theta_0$ be the smoothing length used to measure
$\gamma_g$. These two smoothing scales can be totally different, and we will
find latter that the better signal-to-noise is
optained when they are equal. This discussion is left aside for
now, and we assume in the following that $\theta_0^{\rm CMB}$
and $\theta_0$ are both larger than a few arcmin.
It is therefore reasonable to assume a Gaussian distribution for the variable
$\Vg=[\gammag,\gammag_g]$, and to write
$\Pc_{\rm lens}(\gamma,\gamma_g,\theta,\theta_g)$ as a multivariate
Gaussian distribution,
\begin{equation}
P_{\rm lens}(\gamma,\gamma_g,\theta-\theta_g)= {1\over 2\pi ({\rm
Det}\Mg)^{1/2}} \exp\left[-\Vg\Mg^{-1}\Vg\right],
\label{Plens}
\end{equation}
whose correlation matrix $\Mg$ is given by
\begin{equation}
\Mg=\pmatrix{ \langle\gamma^2\rangle &  \langle \gamma\gamma_g\rangle
\cr \langle\gamma\gamma_g\rangle & \langle\gamma_g^2\rangle}= {1\over
2}\pmatrix{\sigma^2 & r\sigma\sigma_g \cr r\sigma\sigma_g &
\sigma_g^2}.
\label{coreq}
\end{equation}
We used the property $\langle
\gamma_i\gamma_j\rangle=\delta^K_{ij}\langle \kappa^2\rangle/2$ (valid
for the weak lensing limit), where $\sigma^2=\langle\kappa^2\rangle$,
$\sigma_g^2=\langle\kappa_g^2\rangle$. $r=\langle
\kappa\kappa_g\rangle/(\sigma\sigma_g)$ is the correlation coefficient
of the shear amplitude between the CMB and the distant galaxies.  Note
that the probability (\ref{Plens}) only depends on the relative
orientation between $\gammag$ and $\gammag_g$ since we can choose any
origin for the orientations. Figure \ref{corcoef} shows (for LCDM) the
correlation function $r$ for a fixed $\theta_0^{CMB}$ as a function of
$\theta_0$. The moments of the convergence are calculated using the
perturbation theory and the non-linear evolution of the power spectrum
given by \cite{pd96}. It is interesting to note on this figure that the
maximum
correlation is reached when $\theta_0$ and $\theta_0^{CMB}$ are nearly
equal.

With a formal calculator Eq.(\ref{proba_lens}) can be integrated
without difficulties over $\gamma$, $\gamma_g$ and $\theta$  in order
to get the final result:
\begin{equation}
\Pc(\theta_g)\d\theta_g={\d\theta_g\over 2\pi}\left(1+3\sqrt{\pi\over
2}{\langle\kappa\kappa_g\rangle\over
\langle\kappa_g^2\rangle^{1/2}}\cos(\theta_g)\right).
\label{theresult}
\end{equation}
This equation constitutes the main result of this paper. It shows how
much the relative orientation of  the CMB ellipticity with the distant
galaxies deviates from a random distribution, because of the
gravitational lensing effect. 

\begin{figure}[t!]
\centering \psfig{figure=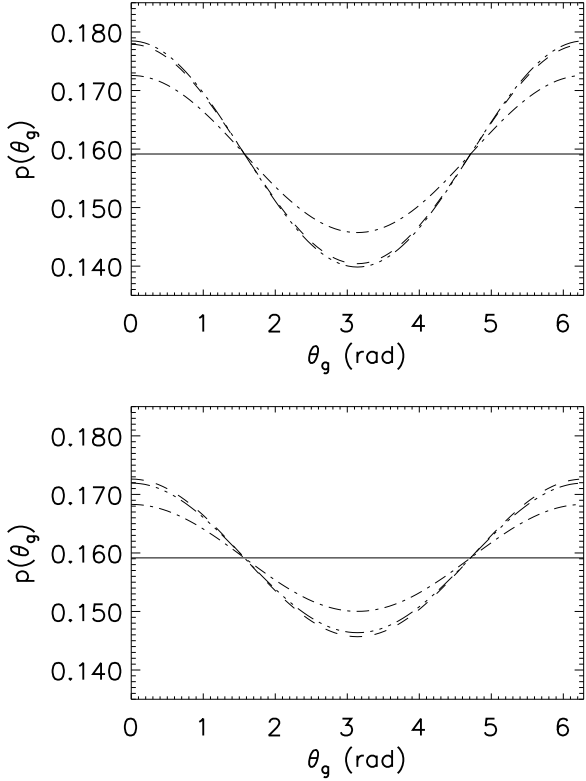,width=0.8\textwidth}
\caption{Probability distribution function of the relative orientation
$\theta_g$ between the CMB ellipticity and the sheared distant
galaxies (Eq.\ref{theresult}).  The beam size is $\theta_0^{\rm
CMB}=2.5'$ for the upper plot and $\theta_0^{\rm CMB}=10'$ for the
lower plot, and the smoothing length of the convergence $\kappa_g$ is
respectively $\theta_0=2'$ and $\theta_0=10'$. The horizontal solid
line represents the uniform distribution of $\theta_g$ in the
un-lensed case. The dot-dashed line is for $\Omega_M=0.3$,
$\Lambda=0$, $\sigma_8=1$, the triple dot-dashed line for
$\Omega_M=0.3$, $\Lambda=0.7$, $\sigma_8=1$ and the dashed line for
$\Omega_M=1$, $\Lambda=0$, $\sigma_8=0.6$.
\label{fig1}}
\end{figure}

Figure \ref{fig1} shows the amplitude of this effect for different
cosmological models and different smoothing scales. We assumed a CDM
power spectrum taking into account its non-linear evolution as
described by \cite{pd96}.  The deviation from a uniform distribution
can be as large as $10\%$, and the effect seems mostly sensitive to
the curvature of the Universe rather than $\Omega_M$ or $\Lambda$.
A possible observable is the average of $\cos(\theta_g)$ over the total
survey area:
\begin{equation}
\langle \cos(\theta_g)\rangle={3\over 2}\sqrt{\pi\over 2}
{\langle\kappa\kappa_g\rangle\over\langle\kappa_g^2\rangle^{1/2}}. 
\label{thequantity}
\end{equation}
If the CMB ellipticities are significantly aligned with the distant
galaxies, then $\langle \cos(\theta_g)\rangle$ should be significantly
larger than zero.  Some values of $\langle \cos(\theta_g)\rangle$ are
given in Table \ref{theestimator} for various cosmological models
and smoothing schemes.

\section{Signal-to-noise analysis}

\subsection{Description}

In order to compute the signal-to-noise of the quantity
(\ref{thequantity}), let us assume that we have at our disposal a CMB
temperature map and a shear map of distant galaxies of angular area
$\Sc$. An estimator $E$ of $\langle \cos(\theta_g)\rangle$ for a
finite number $N$ of measurements at locations $\thetag_i$ is,
\begin{equation}
E={1\over N}{\displaystyle \sum_{i=1}^N \cos\left[\theta_g(\thetag_i)\right]}.
\end{equation}
The ensemble average of the variance of this estimator is
\begin{eqnarray}
\langle E^2\rangle&=&{\langle \cos^2(\theta_g)\rangle\over
N}+{N-1\over N}\int{{\rm d^2}\thetag_1\over \Sc} \int{{\rm
d^2}\thetag_2\over \Sc}\times\nonumber\\ && \langle
\cos\left[\theta_g(\thetag_1)\right]\cos\left[\theta_g(\thetag_2)\right]
\rangle.
\end{eqnarray}
The first term accounts for the variance due to the random intrinsic
orientation of the ellipticities. The second term is the contribution
of the cosmic variance. $N$ can be either the number of individual
galaxies or the number of cells in which the galaxy ellipticities are
averaged. In any case, for $\Sc$ large enough it is
reasonable to take $N\rightarrow\infty$
\footnote{For example for a $900deg^2$ survey there are $N\simeq
9.10^7$ objects with $I<24~mag$.}, for  which only the cosmic variance
term matters. Therefore the variance of $\langle
\cos(\theta_g)\rangle$ is approximated by

\begin{equation}
\sigma^2_{\langle \cos(\theta_g)\rangle}\simeq\int_\Sc {{\rm
d^2}\thetag_1\over \Sc}{{\rm d^2}\thetag_2\over \Sc} \langle
\cos\left[\theta_g(\thetag_1)\right]\cos\left[\theta_g(\thetag_2)\right]
\rangle.
\label{SsurN_def}
\end{equation}
The correlator $C=\langle
\cos\left[\theta_g(\thetag_1)\right]\cos\left[\theta_g(\thetag_2)\right]
\rangle$ depends on
the temperature correlation function and on the convergence correlation
function, and it has to be calculated from the joint probability
$\Pc(\gammag_g(\thetag_1),\gammag_g(\thetag_2),\hat\eg(\thetag_1),\hat\eg(\thetag_2))$.

\subsection{Calculations}

The calculations are simplified if we assume that the
temperature fluctuations are un-lensed and uncorrelated with the
orientation of the galaxies. This will give, by construction, the
signal-to-noise of the CMB-lens positive 
detection against the hypothesis of no lensing on CMB. In
that case the lensing and the CMB distribution functions are separable and
can be written,
\begin{eqnarray}
\Pc(\gammag_g(\thetag_1),\gammag_g(\thetag_2),\eg(\thetag_1),\eg(\thetag_2))=
&&\nonumber\\
\Pc_{\rm lens} (\gammag_g(\thetag_1),\gammag_g(\thetag_2))\ 
\Pc_{\rm CMB}(\eg(\thetag_1),\eg(\thetag_2)).&&
\label{jointproba}
\end{eqnarray}
Using Eq. (\ref{e_def}), it is then possible to re-express the latter
distribution in terms of the variables $\ggr (\thetag_1)$, $\ggr
(\thetag_2)$, $\tau_1$ and $\tau_2$, taking advantage of the fact that
$\Pc_{\rm CMB}(\ggr (\thetag_1), \ggr (\thetag_2))$ follows a Gaussian
distribution. Therefore $\Pc_{\rm CMB}(\ggr (\thetag_1), \ggr
(\thetag_2))$ has the same form as in Eq.(\ref{Plens}) with

\begin{eqnarray}
\Vg&=&[\ggr(\thetag_1),\ggr(\thetag_2)]\nonumber\\ \Mg&=&{1\over
2}\pmatrix{ \langle\tau^2\rangle & \langle
\tau(\thetag_1)\tau(\thetag_2)\rangle \cr \langle \tau(\thetag_1)
\tau(\thetag_2) \rangle & \langle\tau^2\rangle},
\label{Plens_noise}
\end{eqnarray}
here the correlation coefficient is defined as the
temperature fluctuations taken at two different locations $\thetag_1$
and $\thetag_2$. Similarly, $\Pc_{\rm
lens}(\gammag_g(\thetag_1),\gammag_g(\thetag_2))$ is given by
(\ref{Plens}) with 

\begin{eqnarray}
\Vg&=&[\gammag_g(\thetag_1),\gammag_g(\thetag_2)]\nonumber\\
\Mg&=&{1\over 2}\pmatrix{ \langle\kappa_g^2\rangle & \langle
\kappa_g(\thetag_1)\kappa_g(\thetag_2)\rangle \cr \langle
\kappa_g(\thetag_1)  \kappa_g(\thetag_2) \rangle &
\langle\kappa_g^2\rangle}.
\label{Pcmb_noise}
\end{eqnarray}
We are now in position to calculate $C$ in which
we rewrite $\cos[\theta_g(\thetag_1)]$ (resp. $\cos[\theta_g(\thetag_2)]$)
as $\cos[\theta_\eg(\thetag_1)-\theta_{\rm gal}(\thetag_1)]$
(resp. $\cos[\theta_\eg(\thetag_2)-\theta_{\rm gal}(\thetag_2)]$), where
$\theta_\eg(\thetag_1)$ is the orientation of the CMB ellipticity at
location $\thetag_1$, and $\theta_{\rm gal}(\thetag_1)$ the orientation
of a galaxy at location $\thetag_1$. It is usefull to define
$\phi_\eg=\theta_\eg(\thetag_1)-\theta_\eg(\thetag_2)$
and $\phi_{\rm gal}=
\theta_{\rm gal}(\thetag_1)-\theta_{\rm gal}(\thetag_2)$, and to express
$\Pc_{\rm lens}(\gammag_g(\thetag_1),\gammag_g(\thetag_2))$
and $\Pc_{\rm CMB}(\ggr (\thetag_1), \ggr (\thetag_2))$ as
a function of $\cos(\phi_{\rm gal})$ and $\cos(\phi_\eg)$ respectively.
Then the only non-vanishing term in the correlator $C$ is

\begin{equation}
C={1\over 2} \langle \cos(\phi_\eg)\cos(\phi_{\rm gal})\rangle.
\label{corsimple}
\end{equation}
The calculation of $C$ makes intervene the
normalized correlation functions $c_\kappa=\langle
\kappa(\thetag_1)\kappa(\thetag_2)\rangle/\langle\kappa^2\rangle$ and
$c_\tau=\langle \tau(\thetag_1)\tau(\thetag_2)\rangle/\langle
\tau^2\rangle$.
%These correlation coefficients are much smaller than
%unity for a survey larger than $100$ square degrees, therefore
%Eq.(\ref{jointproba}) is expandable in powers of $c_\kappa$ and
%$c_\tau$.
It turns out that the calculation of (\ref{corsimple})
is tractable analytically if we assume that $c_\kappa$ and $c_\tau$
are small compared to unity (then we can expand the distribution function
with respect to $c_\kappa$ and $c_\tau$). This is always a well justified
approximation
for large surveys as those discussed here (for which the $c$
coefficients are at percent level). After a straightforward manipulations
(that can easily be done with a formal calculator) we find
\begin{eqnarray}
&&\langle \cos[\theta_g(\thetag_1)]\cos[\theta_g(\thetag_2)]\rangle
\simeq\nonumber\\
&&{1\over 2}\left({c_\tau^2\over 2}+{7c_\tau^4\over 48}+...\right)
{\pi\over 32}\left(8c_\kappa+
c_\kappa^3+...\right).
\end{eqnarray}
The leading order in the correlation coefficient 
$c_\tau$ and $c_\kappa$ is thus:
\begin{eqnarray}
&&
\langle \cos\left[\theta_g(\thetag_1)\right]
\cos\left[\theta_g(\thetag_2)\right]\rangle=\nonumber\\
&&{\pi\over 16} \left[{\langle\tau(\thetag_1)\tau(\thetag_2)\rangle
\over \langle \tau^2\rangle}\right]^2 
{\langle\kappa(\thetag_1)\kappa(\thetag_2)\rangle\over \langle
\kappa^2\rangle},
\label{finalnoise}
\end{eqnarray}
which has to be integrated over the survey size
in order to give the desired result (\ref{SsurN_def}). 
\subsection{Temperature and convergence correlation functions}

In order to estimate (\ref{finalnoise}) one needs the temperature
and the convergence correlation functions.

The temperature correlation function can be evaluated from the
angular CMB power spectrum $\Cc_l$ given by the CMBFAST code
(Seljak \& Zaldarriaga 1996). However, we have to include the
beam size effect (which tends to enlarge the temperature correlation function
and therefore to increase the noise) and to correct for the instrumental
noise. This can be achieved by a suitable Wiener filtering of the power
spectrum. The filtered $\tilde\Cc_l$ are given by

\begin{equation}
\tilde\Cc_l={\Cc_l {\rm e}^{-l^2\sigma_b^2}\over \Cc_l {\rm
e}^{-l^2\sigma_b^2}
+\Cc_{\rm noise}}\Cc_l {\rm e}^{-l^2\sigma_b^2},
\end{equation}
where $\sigma_b=\theta^{CMB}_0/\sqrt{8\log(2)}$ gives the beam size and
$\Cc_{\rm noise}=\sigma_{\rm pix}^2\Omega_{\rm pix}
\simeq 2.10^{-17}$ is the typical level
of noise for experiments like MAP and PLANCK ($\sigma_{\rm pix}^2$ is the
noise variance per pixel, and $\Omega_{\rm pix}$ the pixel solid angle).
The correlation function is 
then given by the standard Legendre Polynomial expansion:

\begin{equation}
\langle\tau(\thetag_1)\tau(\thetag_2)\rangle={1\over 2\pi} {\displaystyle
\sum_l} \tilde\Cc_l(l+{1\over 2})P_l(\cos(\alpha)),
\end{equation}
where $\alpha$ is the angle between $\thetag_1$ and $\thetag_2$.

The convergence correlation function is calculated as in
Section 3, using the
perturbation theory with the non-linear power spectrum
evolution as described in \cite{pd96}.

\subsection{Results}

We then performed a numerical integration of (\ref{finalnoise}) for
standard-CMD, open-CDM and $\Lambda$-CDM cosmological models. 
The signal-to-noise
is given in the Table \ref{theestimator} for different cosmologies,
smoothing lengths. The survey size is assumed to be
$900~{\rm deg}^2$. The signal to noise ratio scales roughly
like the square root of the area for larger surveys. A very good
signal-to-noise ratio ($\sim 9$) can be obtained at small smoothing
scale, however even in the large smoothing scale case a significant detection
can be obtained. This result contrasts with previous analysis of lensing
on CMB where these typical signal-to-noise ratios were obtained with
a whole sky survey.

\begin{figure}[t!]
\centering \psfig{figure=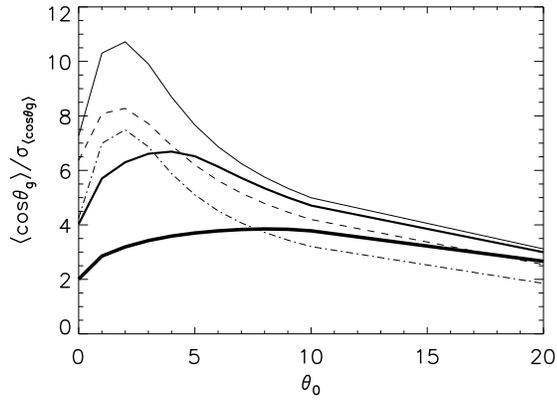,width=0.45\textwidth}
\caption{Signal-to-noise ratio of $\langle \cos(\theta_g)\rangle$ as a
function of the convergence smoothing scale $\theta_0$. The CMB and
the galaxy surveys have both a size of $900~deg^2$. 
The cosmological
models are standard-CDM (dashed line), open-CDM (dotted line) and
$\Lambda$-CDM (solid lines). The beam size is always 2.5'
except for the thick line ($\Lambda$-CDM, beam=5') and
thick thick line ($\Lambda$-CDM, beam=10').
\label{ssurnfunc}}
\end{figure}

Figure \ref{ssurnfunc} shows how the signal-to-noise evolves
with $\theta_0$ for different values of the beam size $\theta_0^{\rm
CMB}$. The maximum signal-to-noise is obtained for $\theta_0$
comparable to $\theta_0^{\rm CMB}$, which is a consequence that at
small scales the sheared galaxies behave essentially like a noise for
the lensed temperature fluctuations, and at large scale the signal
is smoothed away.

\section{Effects of secondary anisotropies}

Undoubtedly, secondary anisotropies and foregrounds can affect our
conclusions and should be examined in details. They may affect our
calculations in three different ways: first some CMB fluctuations might
be generated at low redshift and the assumption the source plane is
located at  $z_{source}\simeq 1000$ breaks down; secondary
anisotropies are correlated with low redshift mass concentrations which might
introduce spurious correlations between CMB ellipticities and
galaxy shear; galactic foregrounds can change our estimation of the
cosmic variance. We compare in the next paragraphs the importance of
various secondary effects.

%Secondary anisotropies can potentially cause a problem for the effect discussed here.
%There are three sources of problem: first some CMB fluctuation might be generated at
%low redshift and the assumption that $z_{source}\simeq 1000$ breaks down. But also, the
%secondary anisotropies are correlated with the low-redshift Universe, which is the source
%of these anisotropies. Therefore the CMB ellipticity $\eg$ can possibliy be correlated
%with the shear of distant galaxies without any lensing effect. The last problem concerns the
%galactic foregrounds, which produces fluctuations uncorrelated with the shear of the nearby
%galaxies, which dilutes our signal.

\subsection{Thermal Sunyaev-Zeldovich (TSZ)}

TSZ effect is the most important source of secondary anisotropies.
It corresponds to the scattering of CMB photons on electrons of the hot
cluster gaz. During the scattering, the photon energy is redistributed
from the low frequency to high frequency part of the black body energy
distribution. The net effect is a deficit of photons at low frequency
(resp. an excess at high frequency), which
creates a negative (resp. positive) fluctuation in the CMB map.
Such fluctuations, essentially centered on clusters of galaxies (where the
gaz is hot enough)
creates a specific ellipticity pattern, correlated with the gravitational
shear generated by those clusters. Therefore the ensemble average $\langle
\cos(\theta_g)\rangle$ (Eq.(\ref{thequantity}))
is not expected to vanish, even in the absence of gravitational
lensing. The importance of this contribution depends on the amplitude and the
extension of TSZ, and it is beyond the scope of this paper to perform an
exact calculation. However we know that the amplitude of
TSZ is proportional to the gaz temperature, which is important in the inner part of
galaxy clusters only, at scales 1-2' (see \cite{perci}, where the TSZ is calculated).
TSZ coming from larger scales (filaments for instance) do not have a significant
contribution (\cite{teg}).
Therefore by observing the CMB at a resolution of 5' or larger, TSZ
should not be a problem. For smaller scales it can become difficult to
disentangle between the lensing effect and the ellipticity-shear correlation
generated by TSZ. However we should emphasize that it is always possible
to suppress this
effect by either observing at 217 GHz (for which TSZ vanishes) or by
cleaning TSZ taking advantage of its specific spectral signature.

\subsection{Kinetic Sunyaev-Zeldovich (KSZ) and Ostriker \& Vishniac (OV) effects}

Although these effects are sometime discussed separately in the litterature,
they have the same physical origin: this is a Doppler effect caused by the
electron bulk motion. OV is a perturbative effect, while KSZ is a
non-linear effect which can be important in the inner part of the clusters.
OV is negligeable at scale larger than 3', but its contribution increases quickly
with decreasing scale, and equals the primary spectrum at 2' (\cite{hw}).
KSZ provides a significant increase of the OV only for scales
smaller than 2' (\cite{hu}). Therefore as long as we work at scales larger 
than 2-3',, KSZ and OV are not important for our purpose. At smaller scales it become
essential to estimate their contribution to $\langle \cos(\theta_g)\rangle$.

\subsection{Integrated Sachs-Wolfe (ISW) and Rees-Sciama effects (RS)}

ISW and RS effects describe the frequency shift of a CMB photon when it
crosses a time evolving gravitational potential well. The induced
fluctuations are pure gravitational effects.
The first order ISW occurs only for Universes different than
$(\Omega=1, \Lambda=0)$ because for the flat $\Omega=1$ Universe the growth
rate of structures is compensated by the Universe expansion. However there
is always a non-linear ISW due to the time evolving potential of non-linear
collapsing structures, which occurs in all cosmologies (the Rees-Sciama
effect). The amplitude of RS was computed by \cite{sel} who
found a very small contribution for $l<4000$. It is therefore irrelevant for our purpose.

\subsection{Galactic foregrounds and point sources}

\cite{teg} calculated the power spectrum contribution of galactic foreground
and extra-galactic point sources. The galactic foreground contamination is
dominant only at very large scales ($l<10$), but the point
sources spectrum might become important for $l>1000$. Fortunately
the clustering properties of the point sources does not play a significant role
(\cite{tof}), therefore their contribution is essentially white noise,
which makes their removal possible, as long as we have a good model for them
(\cite{guider}).

\section{Conclusion}

We have computed the amplitude of the correlation between
the lensed CMB ellipticities and the lensed galaxy ellipticities to the
leading order of perturbation theory.
We found that the modulation of the relative angle 
distribution can be as high as 10\% for a smoothing scale of $2.5'$ and
$5\%$ for $10'$. This lensing effect extends to scales well beyond the
anisotropies generated by the foreground contamination and therefore
it should be easily detectable with a low resolution experiment 
provided that the survey is large enough.
In particular, future wide galaxy surveys (SLOAN) used jointly with
all sky CMB experiments
(MAP, PLANCK) are very promising for the measurement of such a lens
effect. A sky coverage of only $30\%$
will improve the signal-to-noise by a factor of 4 those given in Table
\ref{theestimator}.
At scales larger than a few arcmin, any detected correlation between the
CMB ellipticities and the shear
of the distant galaxies can only be produced by lensing (since foregrounds
contribution is too weak), this does not depend whether the CMB is intrinsically
Gaussian or not (although we used the Gaussian approximation for the calculations).

At the arcmin scale and below, the secondary fluctuations becomes important, and
the contribution of each foreground should be carrefully estimated.
Moreover, at such small scales, the CMB and shear fields cannot be approximated
by Gaussian field, and one might go beyond by considering a weakly non-Gaussian
distribution like the Edgeworth expansion.

The computation of the signal to noise ratio were made in a simplified
way. In particular we assumed that the smoothing $\theta_0^{\rm CMB}$ for
$\kappa$ takes most of the effects associated with the beam smoothing,
which we believe should be a good approximation as long as the
ellipticity orientations only are concerned.
We plan to check this aspect in numerical simulations.

\acknowledgments
We thank Simon Prunet and Dmitry Pogosyan for usefull discussions
on this subject and for a carefull reading of the manuscript.
We all thank IAP for hospitality and the Terapix data center
(http://terapix.iap.fr/) for 
computing  facilities. F.B. and K.B. thank CITA for hospitality.
We thank M. Zaldarriaga and U. Seljak for the use of their code CMBFAST.

%\clearpage

\begin{deluxetable}{lcccc}
\footnotesize \tablecaption{Values of $\langle cos(\theta_g)\rangle$
and its signal-to-noise \label{theestimator}} \tablewidth{0pt}
\tablehead{ 
Model & $\langle \cos(\theta_g)\rangle$ for $\theta_0=5'$ & $S/N$ for
$\theta_0=2.5'$ & $S/N$ for $\theta_0=5'$ & $S/N$ for $\theta_0=10'$ } 
\startdata 
standard-CDM ($900~deg^2$)   & 0.057 & 8.3  &  5.4 &  3.3 \\ 
$\Lambda$-CDM ($900~deg^2$)  & 0.054 & 10.7 &  6.7 &  3.9 \\ 
Open-CDM ($900~deg^2$)       & 0.040 & 7.5  &  4.4 &  2.3 \\ 
\enddata
\end{deluxetable}

%\clearpage

\end{document}